\documentclass{article}
\usepackage{dcase2018,amsmath,graphicx,url,times,booktabs,tabularx,amssymb,amsmath,subcaption}
\usepackage{multicol,multirow}

\title{SAMPLE MIXED-BASED DATA AUGMENTATION FOR DOMESTIC AUDIO TAGGING}

\name{
	Shengyun Wei$^{1}$,
	Kele Xu$^{1,2}$,
	Dezhi Wang$^{3}$,
	Feifan Liao$^{1}$,
	Huaimin Wang$^{2}$,
	Qiuqiang Kong$^{4}$,
}
\address{$^1$ National University of Defense Technology, Information and Communication Dept., Wuhan, China,\\
	yumocloud@protonmail.ch, liaofeifan@126.com\\
	$^2$ National University of Defense Technology, Computer Dept., Changsha, China,\\
	kelele.xu@gmail.com, whm\_w@163.com\\
	$^3$ National University of Defense Technology, College of Meteorology and Oceanography, Changsha, China,\\
	wang\_dezhi@hotmail.com\\
	$^4$ University of Surrey, Center for Vision, Speech and Signal Processing, Guildford, UK, \\
	q.kong@surrey.ac.uk\\
}

\begin{document}

\ninept
\maketitle

\begin{sloppy}

\begin{abstract}
Audio tagging has attracted increasing attention since last decade and has various potential applications in many fields. The objective of audio tagging is to predict the labels of an audio clip. Recently deep learning methods have been applied to audio tagging and have achieved state-of-the-art performance, which provides a poor generalization ability on new data. However due to the limited size of audio tagging data such as DCASE data, the trained models tend to result in overfitting of the network. Previous data augmentation methods such as pitch shifting, time stretching and adding back-ground noise do not show much improvement in audio tagging. In this paper, we explore the sample mixed data augmentation for the domestic audio tagging task, including mixup, SamplePairing and extrapolation. We apply a convolutional recurrent neural network (CRNN) with attention module with log-scaled mel spectrum as a baseline system. In our experiments, we achieve an state-of-the-art of equal error rate (EER) of 0.10 on DCASE 2016 task4 dataset with mixup approach, outperforming the baseline system without data augmentation.
\end{abstract}


\section{Introduction}
\label{sec:intro}

Audio tagging is to label each audio recording with one or more of a multi-label set of labels. This task has many applications such as audio surveillance \cite{Souli2018Audio}, recommendation system \cite{Cano2005Content} and animal populations monitoring \cite{Sprengel2016Audio}, where determining the presence of events in the acoustic scene is the top priority. Manually tagging audio clips is time-consuming and tedious process with growing amounts of data. Consequently, several audio tagging challenges such as DCASE 2016-2017 \cite{dcase2016web}, \cite{dcase2017web} have been held in recent years. Deep convolutional recurrent neural network with attention module have achieved the state-of-the-art performance on DCASE 2016 dataset. Due to the size of many audio tagging datasets are limited to hours \cite{mesaros2016tut}, it remains as a challenge to improve the generalization ability of the network, especially when the training data size is limited. Selection of the model complexity is important for a deep neural network to obtain better generalization performance. For example, by increasing the complexity of a model, the representational ability can be increased, however, it also might increase the possibility of overfitting \cite{Hawkins2004The}.

To increase the generalization ability of the deep neural networks, sustainable efforts have been proposed. For example, dropout \cite{Srivastava2014Dropout} and batch normalization \cite{Ioffe2015Batch} are widely-used regularization techniques for the hidden states of the network. To regularize the intermediate layers in a neural network, several variants have been proposed, such as max-drop and stochastic dropout \cite{Park2016Analysis}. Moreover, shake-shake regularization \cite{Gastaldi2017Shake} and shake drop regularization decrease error rates by disturbing learning \cite{Yamada2018ShakeDrop}.

On the other hand, data augmentation is a crucial component of the state-of-the-art methods\cite{Simard2000Transformation} for different tasks. For example, Random flipping, random cropping and horizontal flipping are widely-used augmentation approaches for the images. For acoustic modeling, pitch shifting, time stretching, and dynamic range compression extend an audio training set with perturbed samples \cite{Salamon2017Deep}. However, these simple data augmentation have a negligible impact on the acoustic modeling performance. In this paper, we propose to use sample mixed data augmentation for audio tagging.

An alternative method for data augmentation is to combine the training samples together, which is called sample mixed-based data augmentation in this paper. For image, SamplePairing synthesizes a new sample from one image by overlaying another image randomly chosen from the training data \cite{Inoue2018Data}. The label of the mixed sample is the same as the label of the first image, which is considered as label-preserving. Compared with label-preserving methods, mixup provides a better generalization ability for image by mix multiple examples, and mixup has demonstrated surprisingly effectiveness \cite{Zhang2018mixup}. In this paper, we target to explore the mixup, SamplePairing and extrapolation data augmentation for audio tagging task in DCASE 2016 (Task 4).

This paper is organized as follows. In Section 2, the preprocessing method is given, while Section 3 describes the data augmentation methods, the employed data augmentation methods include mixup and its variants, SamplePairing and sample extrapolation. Section 4 provides a detailed description of our network architecture, and Section 5 gives the experimental results evaluated on the development dataset. Finally, Section 6 summarizes the paper and provides conclusions.

\section{CONVOLUTIONAL RECURRENT NEURAL NETWORK WITH ATTENTION AND LOCALIZATION}
\label{sec:crnn}

\begin{figure*}[t]
	\centering
	\includegraphics[width=\textwidth,trim=1 1 1 1,clip]{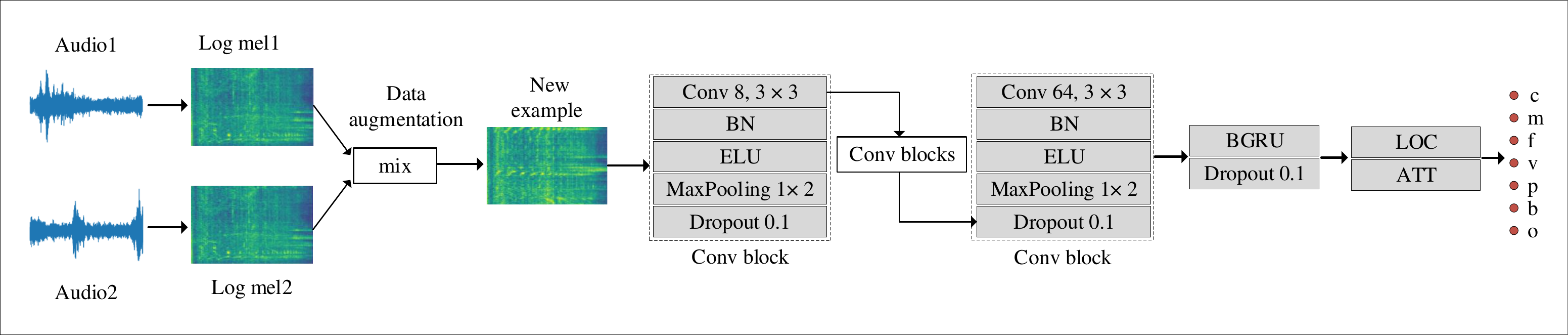}
	\caption{System architecture for audio tagging. Input is log-scaled mel spectrogram with 124 frames and 128 mel frequency bins. Data augmentation is operated on the input samples. There are 7 conv blocks and the number of filters in the convolution layers is 8, 16, 32, 64, 64, 64, 64 respectively.}
	\label{fig:system}
\end{figure*}

Deep convolutional neural networks(CNNs) can provide superior performance for audio tagging \cite{choi2016automatic},  \cite{adavanne2017sound}. For acoustic models, CNNs have been used extensively compared to fully-connected DNNs. Due to modeling local correlations with CNNs can capture spatial local correlation information effectively. Moreover, CNNs reduce spectral variations within audio signals by replicating weights across time and frequency.

Fig.1 shows the architectures of employed CRNN for the audio tagging task. For our neural network architecture, three main parts are employed: time-data augmentation for the frequency representation in the sample space, feature extraction by CRNN and classification to generate output. Log mel spectrograms with 128 frequency bins are used as the input representation. In the Fourier transform, a Hamming window with size of 1024 is used. A stack of convolutional layers are used to extract robust features. All convolutional layers with small kernel size of 3 $\times$ 3 are followed by a batch normalization layer, a max-pooling layer with size 1 $\times$ 2. A dropout layer with ratio 0.1 is used for prevent overfitting. Each block is called Conv block for short. Exponential linear units (ELUs) is used, due to ELUs lead to faster learning and to better generalization performance than ReLUs \cite{Clevert2015Fast}. In addition, Binary cross-entropy is used as the loss function, which gives better results than the quadratic cost \cite{nielsen2015neural}.

Attention and localization based deep convolutional recurrent model (ATT-LOC) achieved the state-of-the-art performance on the evaluation set \cite{Xu2017Attention}. Our structure is similar to ATT-LOC, as both including a deep convolutional recurrent model with an attention module and a localization module. The attention mechanism can effectively reduce the impact of background noise on the output and make the classifier to pay more attention to the frame in which the acoustic events occurr. The localization module detects the onset and offset time of acoustic events in the audio chunk. More details of the attention and localization mechanism could be found in \cite{Xu2017Attention}.

However, some modifications have been made based on the ATT-LOC and the differences between our model and ATT-LOC model include: 1) Different input. The basic features along with the spatial features are concatenated to be fed into the model in ATT-LOC, whereas we just use the basic features of mono audio. This is due to using spatial information of stereo audio to improve results in audio tagging might not always work. The ATT-LOC receives no benefit at all from interaural phase differences and interaural magnitude differences, while only IMD has benefits \cite{Xu2017Attention}. It all depends on the context in which sound events occur and their relative location. Furthermore, mono audio is more general and accessible. 2) Different convolutional layers. ATT-LOC has only one convolutional layer with big kernel size of 30 $\times$ 1, On the contrary, inspired by VGG \cite{Simonyan2014Very}, our network have 7 convolutional layers which are equipped with small kernel size of 3 $\times$ 3) In addition, the number of convolutional layers can be changed automatically according to the input. Due to the CNN can extract the features of different levels, the more layers the network has, the richer features of different levels can be extracted. Moreover, the more abstract the features of network extraction are, the more semantic information there is. However, simply increasing the depth can result in gradient dispersion or gradient explosion. In fact, we have tried to use deeper networks, such as VGG, but the results have not been satisfactory. We conjecture that the size of the dataset does not match the model capability.

\section{SAMPLE MIXED-BASED DATA AUGMENTATION}
\label{sec:samplemixed}

\subsection{Mixup}
\label{ssec:subhead}

Data augmentation aim to expand the training data size by creating new samples, with the goal to reduce the generalization gap between the training and test data. Recently, mixup method has provided better performance for many tasks \cite{xu2018mixup}. In more detail, mixup generates synthetic samples using interpolation in a manner, which is not label-preserving. Unlike the previous attempts to encoder the label using the one-hot encoder approach, the new labels for mixed samples do not belong to two classes, but using the weighted probability of the label. From another perspective, mixup calculates the cross-entropy loss on the two labels with the weighted input, and the two final losses are weighted. The training set can be seen as a bunch of scatters distributed in high-dimensional space. Many new data points between the training set scatter are created by mixup. With the expanded dataset, the relative distance between the scatter points is reduced. To be simpler and more efficient, mixup is applied to a single minibatch and its shuffled version. And the new minibatch can be represented by:

\begin{equation}
\label{eqn:mixup1}
\begin{aligned}
x_n &= \lambda *x_i + (1-\lambda)*x_j \\
y_n &= \lambda*y_i+(1-\lambda)*y_j
\end{aligned}
\end{equation}

where mixing proportion $\lambda \sim{} Beta(\alpha,\alpha)$, for $\alpha \in (0, \infty)$, and $\lambda \in [0, 1]$, $(x_i, y_i)$ and $(x_j,y_j)$ are pairs of samples selected from a minibatch.

\subsection{SamplePairing}
\label{ssec:subhead}

Analogously, SamplePairing \cite{Inoue2018Data} synthesize a new sample by taking an average of two inputs vectors. The label of the mixed sample is the same as the first sample. Hence, Sample pairing constructs new training examples as:

\begin{equation}
\label{eqn:Samplepairing}
\begin{split}
x_n &= 0.5 *x_i + 0.5*x_j \\
y_n &= y_i
\end{split}
\end{equation}

Once new samples have been created by interpolating and pairing between two log mel features, they can be used directly as the input for a deep neural network model.

\subsection{Mixup with label preserving(mixup\_{}lp)}
\label{ssec:subhead}

Mixup\_{}lp is a combination of mixup and SamplePairing. New samples are generated by the fashion of linear interpolation, while in this way that does not use convex combinations of labels.

\subsection{Extrapolation}
\label{ssec:subhead}

Extrapolation operator is used to generated useful synthetic examples in feature space \cite{devries2017dataset}. It is worthwhile to note that, sample mixed-based data augmentation can be either interpolation or extrapolation between a pair of samples in input space. Unlike mixup, which only explore the interpolation between two samples, we explore both interpolation and extrapolation exploration for the domestic audio tagging in this paper. In more detail, extrapolation between two samples is a different alternative of linear combination. However, binary cross-entropy error has negative values when generate the labels by extrapolation, due to the categories are labeled as values larger than 1 instead of 0 and 1, which confuses the classifier. So we have:

\begin{equation}
\label{eqn:extra}
\begin{split}
x_n &= (1+\lambda)*x_i - \lambda*x_j \\
y_n &= y_i
\end{split}
\end{equation}

SamplePairing, mixup\_{}lp and extrapolation are sample mixed data augmentation of label preserving, which can used for Semi-supervised Learning.

\section{EXPERIMENTS AND RESULTS}
\label{sec:pagestyle}

\begin{table*}[]

\scriptsize

\centering

\caption{Experimental results on the evaluation set of the DCASE 2016 audio tagging challenge.}

\label{Tab01}

\begin{tabular}{ccccccccccc}

\toprule

\multirow{2}{*}{Model}&\multirow{2}{*}{Date augmentation}&\multicolumn{8}{c}{EER}&\multirow{2}{*}{Var($10^{-3}$)}\\

\cmidrule(r){3-10}

&{}&{c}&{m}&{f}&{v}&{p}&{b}&{o}&{Avg}\\

\midrule

DAE-DNN&$\sim{}$&0.21&0.15&0.21&0.02&0.18&0.01&0.26&0.15&9.45\\
CGRNN&IMD&0.17&0.16&0.18&0.03&0.15&0.00&0.24&0.13&7.39\\
ATT-LOC&IMD&0.09&0.14 &0.17&0.03&0.12&0.01&0.24&0.11&6.36\\
CRNN&mixup($\alpha = 0.0$)&0.09&0.11&0.14&0.04&0.15&0.06&0.26&0.13&6.13\\
CRNN&mixup($\alpha = 0.1$)&0.10&0.23&0.15&0.02&0.15&0.03&0.23&0.13&7.30\\
CRNN&mixup($\alpha = 0.5$)&0.09&0.16&0.11&0.03&0.14&0.03&0.24&0.11&5.56\\
CRNN&mixup($\alpha = 1.0$)&0.09&0.12&0.11&0.02&0.12&0.03&0.26&0.11&6.25\\
CRNN&mixup($\alpha = 1.5$)&0.10&0.14&0.11&0.03&0.10&0.01&0.20&0.10&4.11\\
CRNN&mixup($\alpha = 2.0$)&0.10&0.11&0.11&0.03&0.11&0.00&0.25&0.10&6.28\\
CRNN&mixup($\alpha = 5.0$)&0.13 &0.14&0.08&0.02&0.12&0.00&0.24&0.11&6.52\\
CRNN&SamplePairing&0.10&0.20&0.15&0.01&0.16&0.03&0.24&0.13&7.26\\
CRNN&mixup\_{}lp($\alpha = 1.5$)&0.12 &0.13&0.12&0.02&0.12&0.00&0.25&0.11&6.52\\
CRNN&extrapolation($\alpha = 1.5$)&0.10&0.16&0.13&0.03&0.14&0.02&0.23&0.12&5.43\\

\bottomrule

\end{tabular}

\end{table*}

\begin{figure*}[t]
    \centering
    \includegraphics[width = 0.3\linewidth]{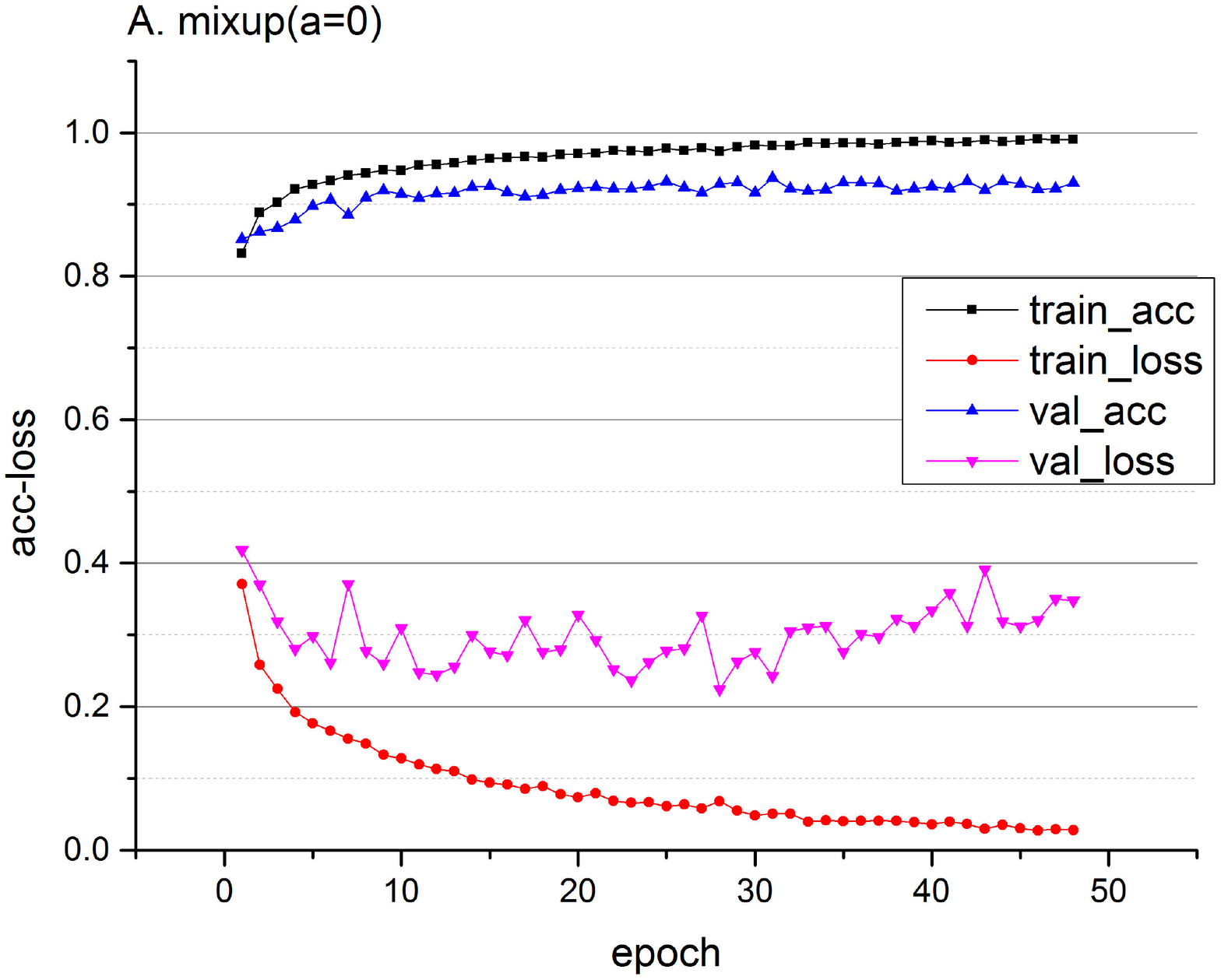}\
    \includegraphics[width = 0.3\linewidth]{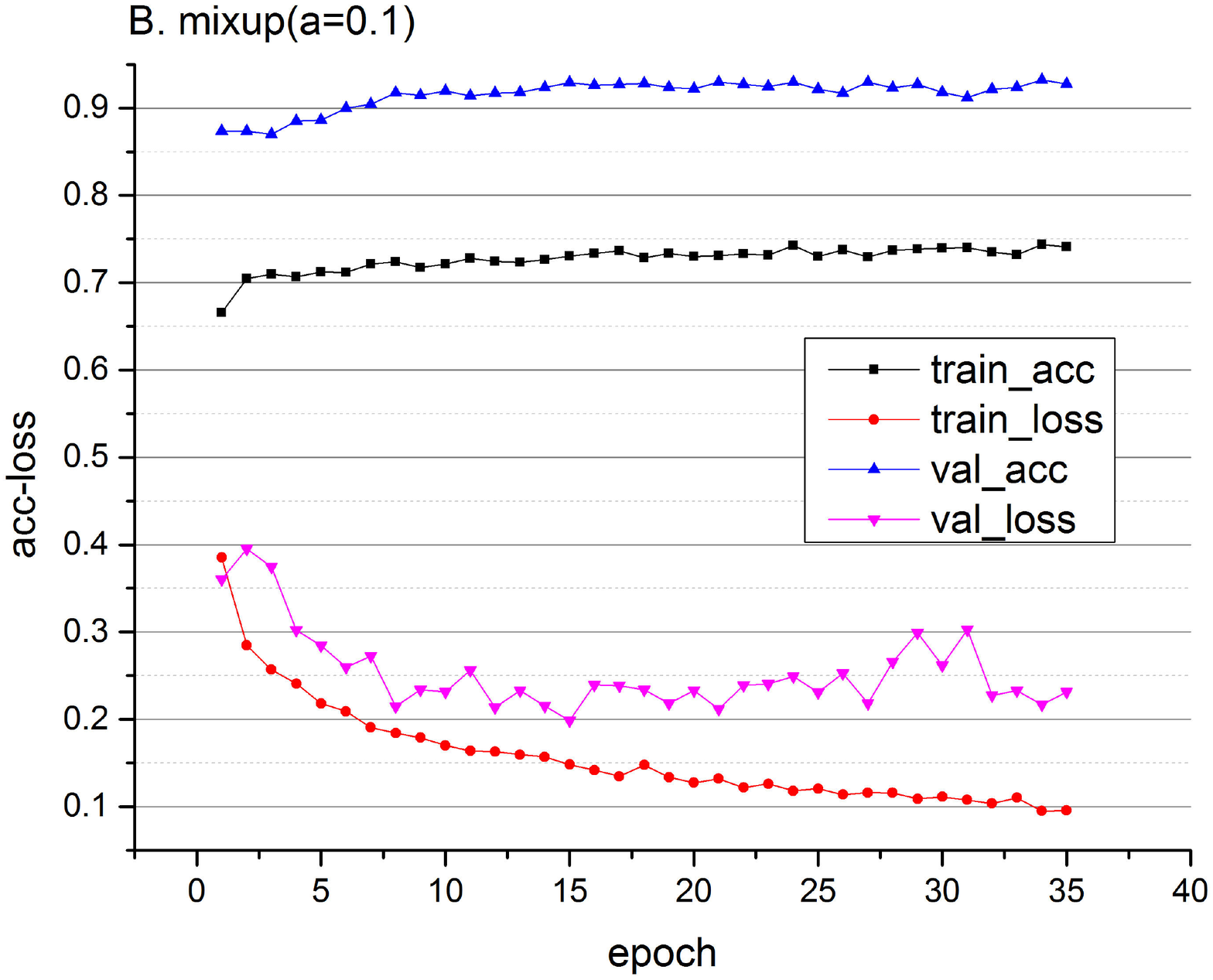}\
    \includegraphics[width = 0.3\linewidth]{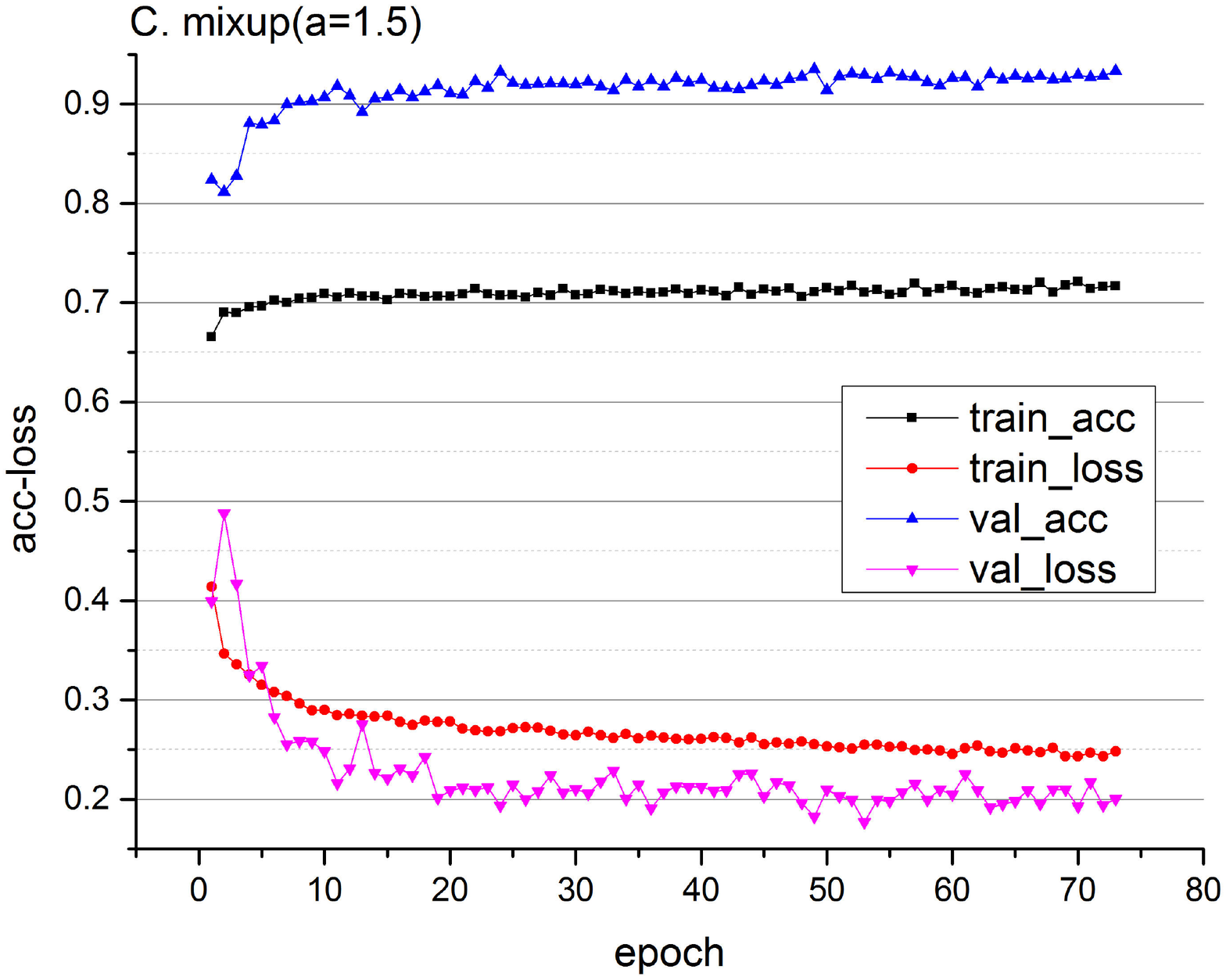}\\[1mm]
    \includegraphics[width = 0.3\linewidth]{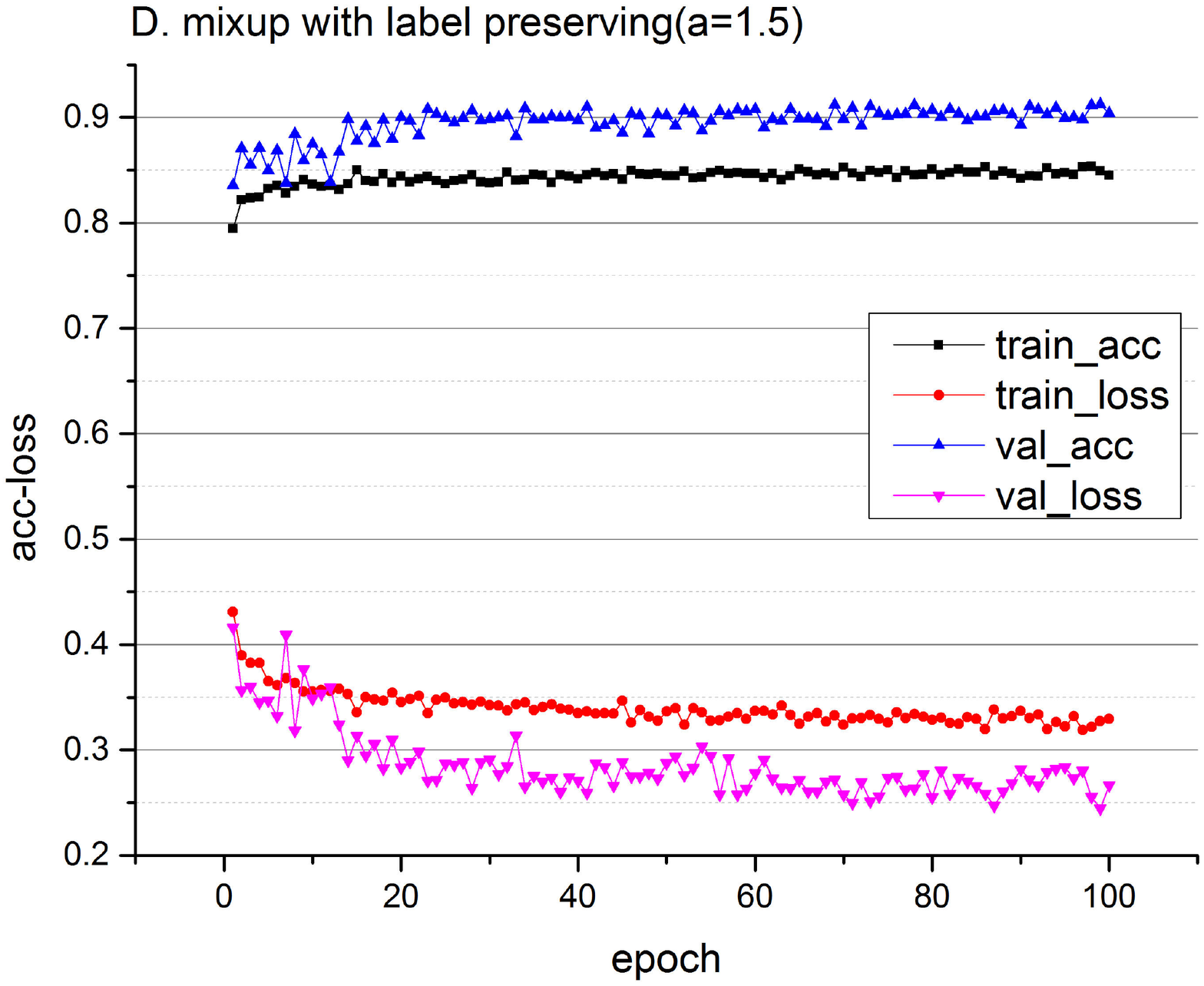}\
    \includegraphics[width = 0.3\linewidth]{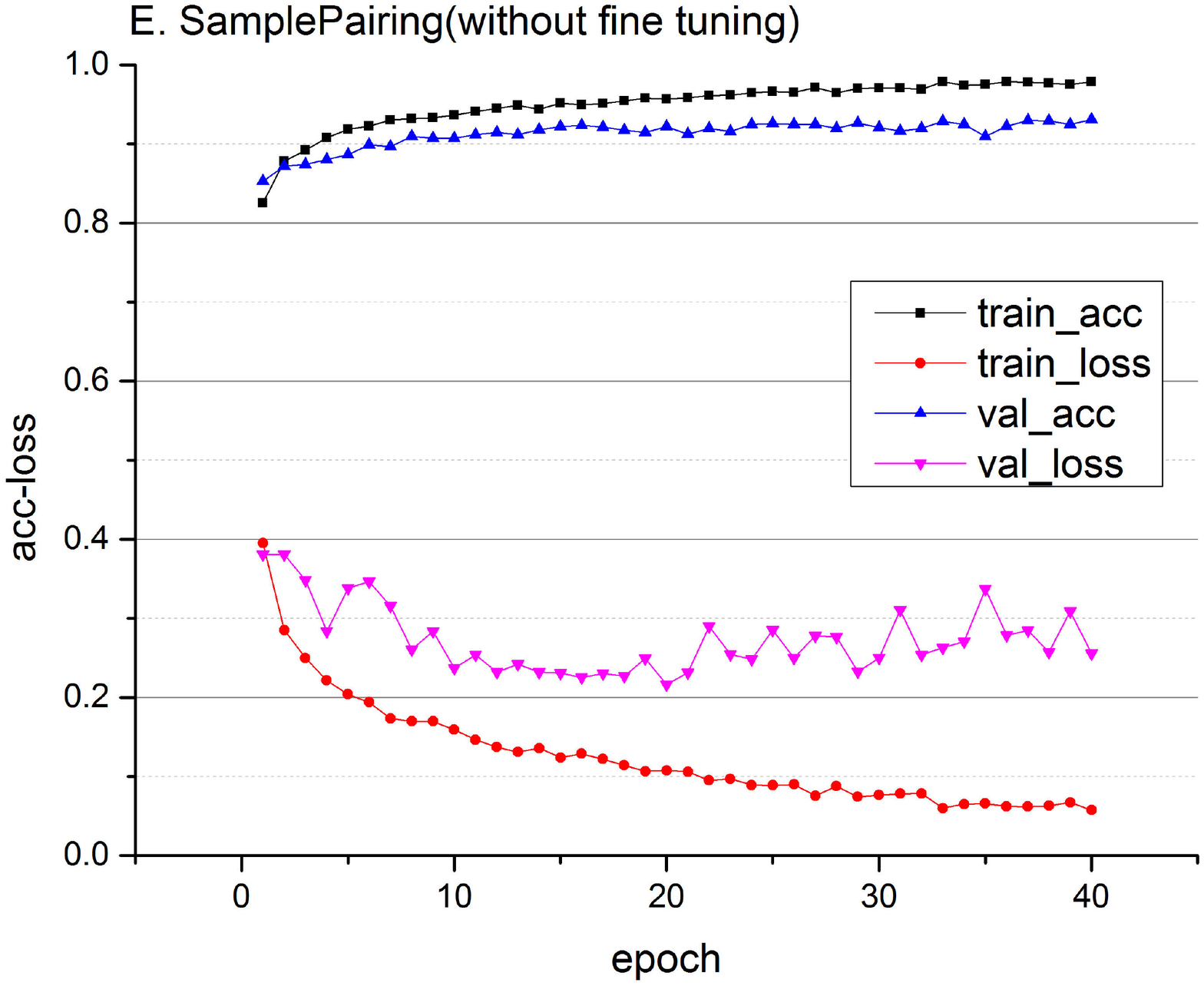}\\[1mm]
    \caption{Model trainning curves using different data augmentation methods.}
    \end{figure*}

\subsection{Dataset and preprocessing}
\label{ssec:subhead}

The proposed methods are evaluated on CHIME-HOME dataset of the DCASE 2016 audio tagging challenge that comprises audio chunks along with corresponding multi-label annotations or ground truth labels \cite{Foster2015Chime}. The annotations are based on a set of 7 label classes, including child speech, adult male speech, adult female speech, video game / TV, percussive sounds, broadband noise and other identifiable sounds, denoted by ‘c’, ‘m’, ‘f’, ‘v’, ‘p’, ‘b’ and ‘o’. Multi-label annotations suggest that the dataset is weakly labeled with chunk level labels rather than event level labels. In fact, an audio chunk may contain multiple sound events without indicating their occurrence time \cite{mesaros2016tut}. The development dataset consists of 1946 4-second chunks with the 16kHz sampling rate in mono. The target is to perform multi-label classification on 4-second audio chunks.

Directly learning features from the raw waveform is subject to the limited size of the training data. Presently, most of the audio tagging systems used frequency-domain features as the input, extracted from the audio signal clip. They are mainly borrowed from the field of speech recognition, such as mel-scale filter banks, log-frequency filter banks and time-frequency filters. In audio analysis tasks, for example, audio scene classification, audio event detection and audio tagging, frequency-domain representation provides superior performance. However, the MFCCs may not maintain locality by the discrete cosine transform projecting the spectral energies into a new basis. As a consequence, the log-mel features computed directly from the mel-frequency spectral coefficients for each frame of raw audio was used as an input of CNN \cite{Piczak2015Environmental}, \cite{Abdel2014Convolutional}. In this paper, we use log-mel features as the input for the neural network.

\subsection{Evaluation and baseline}
\label{ssec:subhead}

The official evaluation method for the challenge is average equal error rate (EER) for five-fold cross-validation. We followed the 5-fold cross-validation setting using the original folds splits. Early stopping is used to monitor the validation loss. Training is interrupted when the validation loss has not improved after 20 epochs. The batch size is set up to 44.

The EER is used as an evaluation metric, which is defined as the error rate at the ROC operating point where the false positive and false-negative rates are equal, and a lower EER represents better system performance.

We apply DAE-DNN \cite{xu2017unsupervised}, CGRNN \cite{Xu2017Convolutional} and ATT-LOC \cite{Xu2017Attention} as the baseline models. CGRNN and ATT-LOC downsamples the stereo audio data from the 48kHz sampling rate into 16kHz. Based on previous experiments, these three models achieved the state-of-the-art performance with 0.15,0.13 and 0.13 EER on the evaluation set, respectively.

\subsection{Results}
\label{ssec:subhead}

Table I shows experimental results of the DCASE 2016 audio tagging challenge, by using different data augmentation approaches. The experiment consists of three parts.

\subsubsection{The effectiveness of the proposed architecture}
\label{sssec:subsubhead}

Firstly, we verify the effectiveness of the proposed novel neural network architecture. without data augmentation, our proposed CRNN model (when $\alpha = 0$) can get acceptable classification performance with 0.13 EER. Incorporating mixup, when $\alpha = 1.5$ and $\alpha = 2$, gain the best performance with 0.10 EER, which is the state-of-the-art performance on the evaluation set of the DCASE 2016 audio tagging challenge.

\subsubsection{Data augmentation of mixed form }
\label{sssec:subsubhead}

we observe that data augmentation of mixed form (including mixup, SamplePairing and extrapolation) can effectively improve the classification. The only exception is the Samplepairing approach without fine tuning,  no significant performance is observed. In more detail, SamplePairing without fine tuning perform poorly to classify the “adult male speech (m)” audio event. In the development dataset, the “adult male speech (m)” event occurs sparsely (number of occurrences is 174) with comparison with other events. New examples generated by fixed interpolation coefficient confused the classifier for minority classes.

\subsubsection{Different ratios for the mixup approach }
\label{sssec:subsubhead}

we explored different ratios for the mixup approach. We choose $\alpha \in\{0, 0.1, 0.5, 1.0, 1.5, 2.0, 5.0\}$. We find that $\alpha \in \{1.5, 2.0\}$ gets the best performance in our experiment. In addition, when $\alpha = 1.5$, the approach has the smallest variance of EER $4.11 \times 10^{-3}$, which indicates that our CRNN model with mixup approach has better stability.

\subsubsection{The model training history }
\label{sssec:subsubhead}

Furthermore, Fig.2 shows the model training history for each epoch, including the training loss and accuracy, as well as the loss and accuracy for the validation dataset. As can be seen from the figure, the results in Fig.2B and Fig.2C get a significantly lower training accuracy (basically stable at around 0.7) and a higher training loss compared with that in Fig.2A (without mixup), whereas the validation loss is the lowest (when $\alpha = 1.5$). In addition, we observe that the bigger $\alpha$, the lower training accuracy will be. When interpolation is not used for labels (as shown in Fig.2D), the training accuracy is obviously better than that in Fig.2C. However, this did not lead to performance improvements. In Fig.2A, Fig.2E and Fig.2F, the gap between training loss and validation loss increases as the training epoch increases. We suppose that this situation may be detrimental to the generalization ability of the model.

\section{CONCLUSION}
\label{sec:CONCLUSION}

In this paper, for the audio tagging task, we use data augmentation of mixed form on the time-frequency representation in input space for training the convolutional recurrent neural network (CRNN). Experiments are conducted on DCASE 2016 (Task 4) task. Based on our experiments, sample-mixed based data augmentation can effectively improve the performance of audio tagging. Moreover, mixup generalizes better than other mixed form data augmentation methods, as it has dramatically decrease the gap between the training and test distribution, which gains the best performance with 0.10 EER, and this is the state-of-the-art performance on the evaluation set of the DCASE 2016 audio tagging challenge.

\bibliographystyle{IEEEtran}
\bibliography{refs}

\end{sloppy}
\end{document}